\documentclass[12pt,twoside]{article}   
\usepackage[super,sort,comma]{natbib}
\usepackage{authblk}

\usepackage{fancyhdr}		

\newcommand{\captionv}[3]{\begin{center}\parbox{#1cm}{\caption[#2]{{\sf #3}}}
        \end{center}}

\usepackage[section]{placeins}   %

\usepackage{graphicx}
\graphicspath{{img/}}
\usepackage{siunitx}
\usepackage{subcaption}

\makeatletter \renewcommand\@biblabel[1]{$^{#1}$} \makeatother
\newcommand{\cen}[1]{\begin{center} #1 \end{center}}

\usepackage{hyperref}
\hypersetup{ colorlinks,
    citecolor=blue,
    filecolor=blue,
    linkcolor=blue,
    urlcolor=blue
}

\usepackage{xcolor}

\definecolor{gray}{rgb}{0.6,0.6,0.6}
\definecolor{red}{rgb}{0.85,0,0}
\definecolor{green}{rgb}{0,0.85,0}
\definecolor{blue}{rgb}{0,0,0.85}
\definecolor{beige}{rgb}{0.92,0.87,0.78}
\usepackage[all]{hypcap}    

\begin{document}





\cen{\sf {\Large {\bfseries Clinical utility of automatic treatment planning for proton therapy of head-and-neck cancer patients using JulianA} \\
\vspace*{10mm}
R. Bellotti$^{1, 2}$, A. Cherchik$^1$, J. Willmann$^3$, A.~Lomax$^{1, 2}$, D. C. Weber$^{1, 4, 5}$, J. Hrbacek$^1$} \\
1. Center for Proton Therapy, Paul Scherrer Institut, 5232 Villigen PSI\\
2. Department of Physics, ETH Zürich, 8092 Zürich\\
3. Accelerator Modelling and Advanced Simulations Group, 5232 Villigen PSI\\
4. Department of Radiation Oncology, University Hospital of Zürich, 8091 Zürich\\
5. Department of Radiation Oncology, Inselspital, Bern University Hospital, University of Bern, 3010 Bern
\vspace{5mm}\\
Version typeset \today\\
}

\pagenumbering{roman}
\setcounter{page}{1}
\pagestyle{plain}
Author to whom correspondence should be addressed. email: jan.hrbacek@psi.ch \\

\clearpage
\begin{abstract}
\noindent {\bf Background:} Automatic treatment planning promises many benefits for both research and clinical environments. For research, autoplanning provides the ability to include treatment planning as a part of a reproducible workflow that can be run at the press of a button after changing any parameter, algorithm or model, which is particularly useful for applications in machine learning and artificial intelligence. For clinics, autoplanning promises to reduce planning time and achieve more comparable treatment plans and thereby reduce inter-planner variability. Further, it can assist clinicians in quality assurance by providing a minimum plan quality standard. Finally, autoplanning is an essential part of patient selection, which is crucial for the advancement of proton therapy itself.\\
{\bf Purpose:} We explore the clinical utility of the recently published open source JulianA software package and its automatic spot weight optimisation in a retrospective planning study. \\
{\bf Methods:} A retrospective planning study using a cohort of $17$ head-and-neck cancer patients treated at our institute. The clinically accepted plans created by dosimetrists (d-plans) were compared to automatically generated JulianA plans (j-plans). The d-plans were created in Eclipse using non-robust optimisation according to our clinical protocol at the time. Both methods used the same beam arrangement. The plans were analysed by two expert reviewers without knowing how each plan was created. They assessed the plan quality and stated a preference.\\
{\bf Results:} All of the j-plans were deemed rather or clearly acceptable, resulting in a higher acceptability than the d-plans. The j-plan was considered superior in $14 (82.4\%)$ cases, of equal quality for $1 (5.9\%)$ and inferior to the d-plan for only $2 (11.8\%)$ of the cases. The reviewers concluded that JulianA achieves more conformal dose distributions for the $15 (88.2\%)$ cases where the j-plans were at least as good as the d-plans. Concretely, the j-plans conform more closely to the prescribed dose levels, achieve better organ-at-risk sparing and deposit less dose to normal tissue than the d-plans by realising a sharper falloff while maintaining similar target coverage.\\
{\bf Conclusions:} The results show that the JulianA is ready to be used as a clinical quality assurance tool and research platform at our institute. While these results are encouraging, further research is needed to reduce the number of spots further and introduce robustness considerations into the optimisation algorithm in order to employ it on a daily basis for patient treatment.\\

\end{abstract}

\newpage     

The table of contents is for drafting and refereeing purposes only. Note
that all links to references, tables and figures can be clicked on and
returned to calling point using cmd[ on a Mac using Preview or some
equivalent on PCs (see View - go to on whatever reader).
\tableofcontents

\newpage

\setlength{\baselineskip}{0.7cm}      

\pagenumbering{arabic}
\setcounter{page}{1}
\pagestyle{fancy}
\section{Introduction}
Treatment planning is an important step in the process of delivering radiotherapy. A good treatment plan fathoms the physical limits of the delivery machine in order to maximise tumour control while minimising side effects. Unfortunately, crafting a high quality treatment plan is a demanding and time consuming task that requires medical physicists or dosimetrists with special training and years of experience~\cite{Coffey2022-aq}.

Automatic treatment planning (autoplanning) promises to shorten the planning time and relieve the dosimetrists from creating treatment plans from scratch, reducing treatment costs and time to treatment. Interest in autoplanning algorithms has substantially increased in recent years \cite{hussein_automation_2018} and it will most probably continue to shape and advance the future of radiotherapy\cite{nystrom_treatment_2020}. The majority of literature focuses on algorithms that belong to the categories of knowledge-based planning (KBP) and multi-criteria optimisation (MCO) and protocol-based automatic iterative optimisation~\cite{hussein_automation_2018}, whereas the latter is not completely automatic and therefore will not be discussed in this study.

KBP leverages a database of previously generated treatment plans for similar patients to predict achievable dose distributions or dose-volume histograms (DVHs)~\cite{hussein_automation_2018, ahervo_artificial_2023}. These predictions are then used as objectives for a so-called mimicking optimisation to obtain deliverable treatment plans. The most prominent example of this category is the commercially available RapidPlan (Varian Medical Systems, Palo Alto, CA). Several studies have shown RapidPlan's utility for photon~\cite{panettieri_development_2019} and proton radiotherapy~\cite{delaney_evaluation_2018, delaney_automated_2018, xu_assessment_2021, van_bruggen_automated_2023} and patient selection for proton treatment~\cite{hytonen_fast_2022}. Other KBP approaches are available in RayStation~\cite{borderiasvillarroel_machine_2023} (RaySearch Laboratories AB, Stockholm, Sweden) or in non-commercial models~\cite{maes_automated_2023}.

MCO algorithms aim to either approximate the entire Pareto front and let a human planner select the final plan (a posteriori MCO) or define the desired point on the Pareto front in advance (a priori MCO). The most prevalent method in this field is Erasmus-iCycle, which has been validated extensively for photon~\cite{Breedveld2012, della_gala_fully_2017, Bijman2021} and proton therapy~\cite{Huiskes2024, Kong2024}.

While MCO methods result in Pareto-optimal plans, they take a long time (\SI{47}{min} for SISS and 2-4h for iCycle~\cite{Kong2024}) to converge. On the other hand, KBP planning takes only minutes, but requires are large dataset of similar cases for training, which is not available at our institute. Further, adjusting a KBP model requires retraining, which can take multiple hours depending on the model and the avaialble hardware. For these reasons, we have developed the JulianA algorithm~\cite{juliana}. JulianA is an priori MCO algorithm based on a weighted sum loss function whose weights are patient-unspecific. However, its objective weights could be fine-tuned for special cases, which is not needed for the patients discussed in this study. JulianA is fast thanks to its use of gradient-based optimisation. Previous research has shown that JulianA is capable of generating reasonable treatment plans for a diverse range of intra- and extracerebral tumours~\cite{juliana}. Contrary to other fast autoplanning solutions such as RapidPlan, JulianA does not require a large number of training patients. For the head-and-neck cases discussed in this study, only two training patients (patient 01 and 02) are needed to tune the parameters of the loss function. The JulianA autoplanning algorithm is implemented in the open source Julia package JulianA.jl~\cite{bellotti_2024_11079457} and freely available under the BSD 3-clause license.

In this study, we conduct a retrospective planning study to explore the clinical utility of JulianA for head-and-neck cases. Such cases provide an excellent test for the power of any autoplanning tool due to the large number of critical organs-at-risk (OARs) such as optical structures, spinal cord, brainstem, pharynx constrictor muscles and diverse salivary glands in close proximity to the target. The envisioned applications of JulianA at this stage are quality assurance (QA) in daily clinical practice and supporting research.

In principle, any autoplanning solution could be used for QA purposes by providing a baseline treatment plan. Warnings would be issued if the plan suggested for treatment is inferior to the baseline plan w.\ r.\ t.\ predefined quality metrics such as target $V_{95\%}$ or maximum, mean dose or any DVH-based values for targets and OARs. The strength of JulianA for such an application is that it creates the autoplan in comparatively little time and is also suitable for clinics with very small datasets. JulianA will be combined with an automatic beam angle selection method called GAMBAS (Geometry-based Automatic Model for Beam Arrangement Selection) that is currently being evaluated at our institute to establish a fully automatic treatment planning pipeline for QA.

JulianA is a comparatively fast autoplanning solution that requires very little to no data for parameter tuning. These properties render it a decent tool to support researchers that typically have neither training nor experience with treatment planning. The use of autoplanning tools such as JulianA relieves researchers from treatment planning, saving time and closing the quality gap between the treatment plans created for research and the treatment plans for patient treatment. For example, machine learning projects for synthetic CT generation (either from magnetic resonance or cone-beam CTs) often need to create treatment plans as part of their validation to analyse the dosimetric impact of the model. Autoplanning simplifies and streamlines the validation of such models.

In this paper, we demonstrate the adequacy of JulianA for QA and research supports by performing a retrospective planning study with two expert reviewers that judge the automatically generated and our dosimetrists's plans and compare them in a blinded way. The emphasis is not on creating plans for treating patients, but to demonstrate that JulianA is able to create high-quality baseline plans for QA purposes and to equip researchers with a reliable, fast and homogeneous way of generating treatment plans of a comparable quality to treatment-approved plans without any human interaction. The goal is therefore not to automatically generate treatment plans of a higher quality, but of a similar quality than the ones created manually in our clinic in a short and practical time.


\section{Methods}
Automatic treatment plans were generated using the open-source automatic planning algorithm JulianA \cite{bellotti_2024_11079457} and two independent expert reviewers, both radation oncologists at our institute, compared them to clinically approved treatment. The inclusion criteria for the cohort were head-and-neck cancer, planned in the Eclipse (Varian Medical Systems, Palo Alto, CA) treatment planning system using non-robust optimisation, including prophylactic lymph-node irradiation and simultaneous integrated boost with dose levels \SI{54.12}{Gy RBE}, \SI{59.40}{Gy RBE}, \SI{69.96}{Gy RBE}. Non-robust optimisation was a requirement because robust optimisation was introduced to our clinic only recently and the number of robustly optimised treatment plans in our clinical database was not deemed to be sufficient for this study. The resulting cohort consists of $17$ patients that were treated at our institute between $2019$ and $2023$. For all of these patients, the CT, structureset and organ-at-risk (OAR) mean and maximum dose constraints as well as the dose distribution that had been accepted for treatment were exported from Eclipse. The dose distribution is called the dosimetrist's plan (d-plan). The previously published JulianA algorithm~\cite{juliana} was used for the autoplanning using the same beam arrangements as the d-plans. Its parameters were tuned based on patients 01 and 02 and used to create a treatment plan (j-plan) for each patient. Then, the d-plan and j-plan were randomly assigned labels A and B and given to the expert reviewers. The expert reviewers were asked to assess the quality as they would during a plan review board meeting. Discussion between the expert reviewers was allowed and only one final rating was given for each plan. The assessment was performed separately for target coverage, OAR sparing, normal tissue sparing and overall plan quality based on a scale from 1-5 (clearly unacceptable, rather unacceptable, unsure, rather acceptable, clearly acceptable). Additionally, a 1-5 preference was given for each of the plan aspects.

The JulianA algorithm was developed at our institute. It is an automatic spot weight optimisation algorithm whose objective weights are not patient-specific. Its input are the CT, structure set and the OAR mean/maximum dose constraints and the beam arrangement, for which we use the same angles as for the corresponding d-plan. The JulianA code was adjusted as follows since the original publication. First, the derivatives of the loss function were hard-coded rather than calculated using automatic differentiation to reduce the memory footprint, which is especially necessary for the big tumours common in head-and-neck patients. Second, a multi-stage optimisation was introduced to accommodate the complexity of head-and-neck cases. The first stage optimises target coverage, the second adds objectives for OAR maximum dose the region of mean-constraint OARs outside of the target in order to prioritise OAR sparing outside of the target. The third stage performs a full optimisation including all the OAR constraints. Finally, the code was heavily refactored to improve user-friendliness.

The j-plans are deliverable and adhere to the minimum deliverable monitor units at our institute to the best of our knowledge, though we have not performed the delivery. In agreement with our clinical protocol during data acquisition, robustness was neither considered for the generation of the d-plans nor for the j-plans. The spot placement and dose calculations for the d-plans were performed in the Eclipse treatment planning system (TPS) because this system is used for head-and-neck patients at our institute. However, the use-case of JulianA would be to be integrated into our in-house planning system FIonA, which uses a different beam model and a less efficient spot placement algorithm than Eclipse. The beam models are different because Eclipse is used for gantry three and FIonA is used for gantry 2. These differences do not limit the meaningfulness of this study: The goal of this study is not to prove JulianA's applicability for patient treatment, but to show that JulianA creates meaningful treatment plans in a short time and without human interaction.


\section{Results}
\subsection{Case studies}
\begin{figure}
    \centering
    \begin{subfigure}{0.75\textwidth}
        \includegraphics[width=\textwidth]{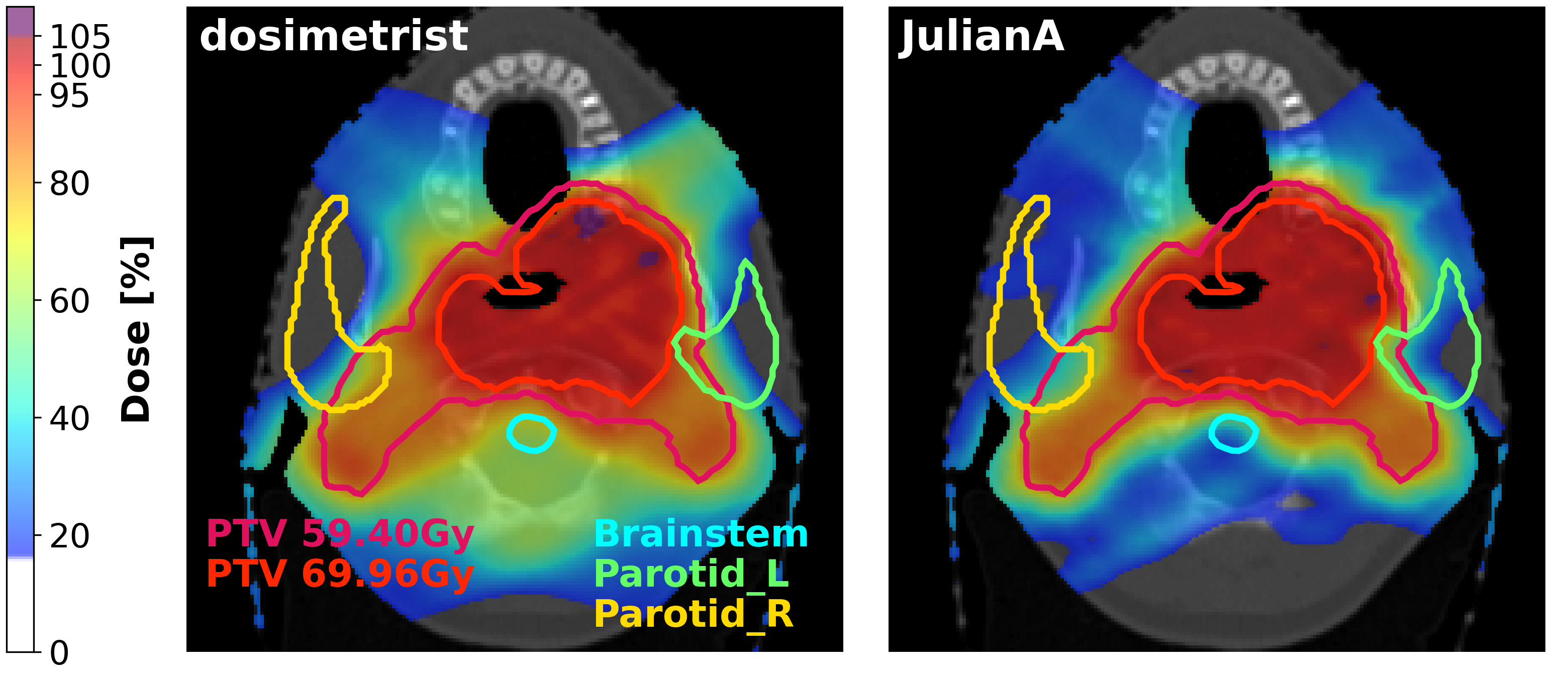}
        \caption{Plan comparison for patient 05. The j-plan (right) exhibits better OAR sparing (i.\ e.\ brainstem, spinal cord and left parotid gland) at comparable target coverage and is therefore preferred by the expert reviewers.}
        \label{fig:case_05}
    \end{subfigure}
    \begin{subfigure}{0.75\textwidth}
        \includegraphics[width=\textwidth]{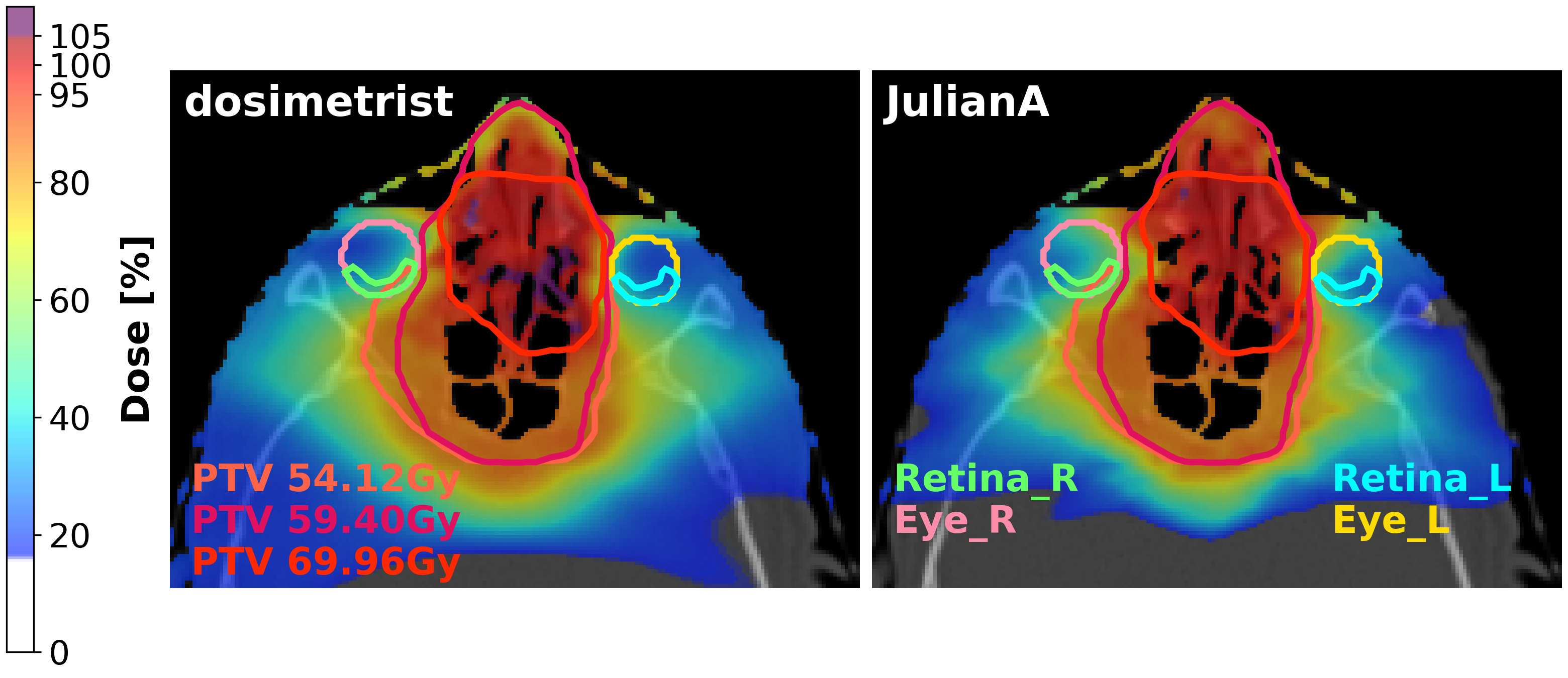}
        \caption{Plan comparison for patient 12. Both dose distributions look similar and are considered equivalent by the expert reviewers.}
        \label{fig:case_09}
    \end{subfigure}
    \begin{subfigure}{0.75\textwidth}
        \includegraphics[width=\textwidth]{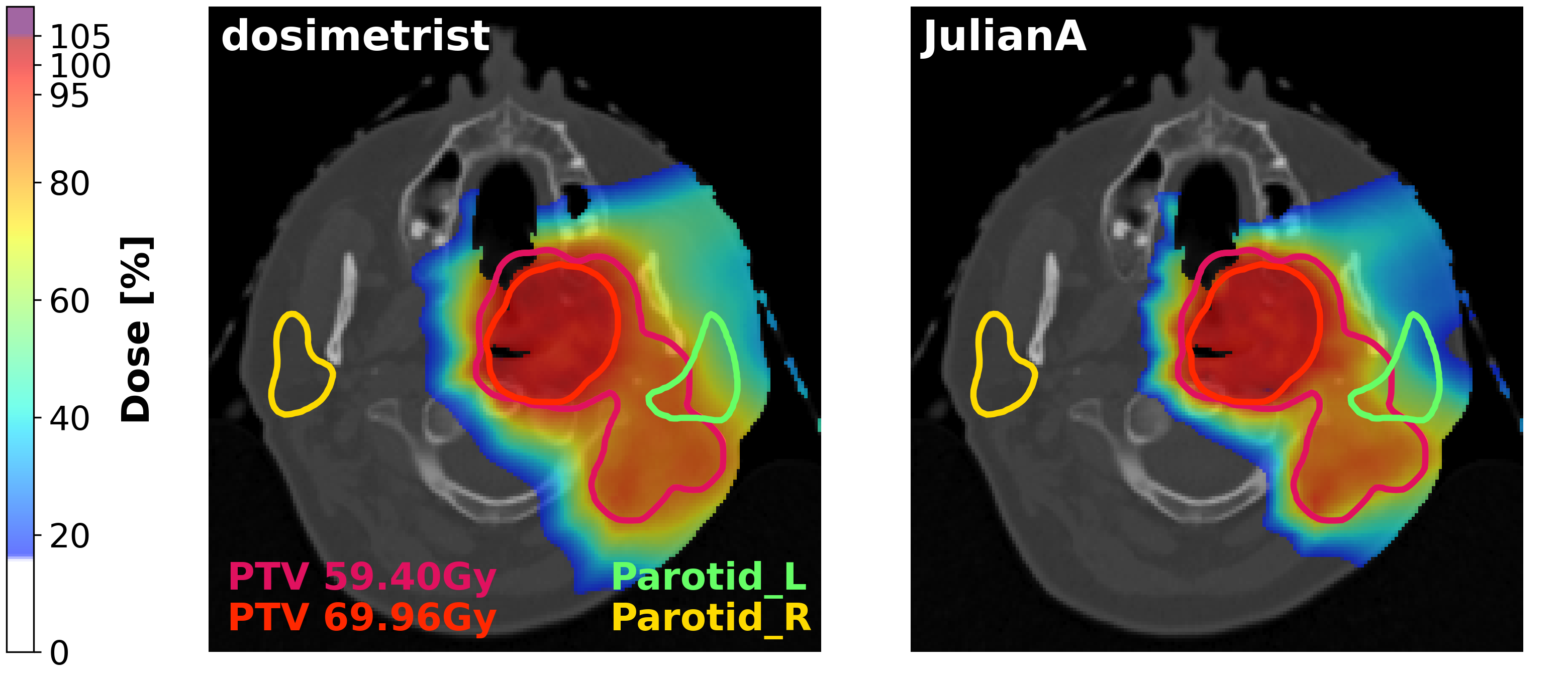}
        \caption{Plan comparison for patient 09. The d-plan improves target coverage by allowing higher dose values to the ipsilateral parotid, which renders it preferable according to the expert reviewers.}
        \label{fig:case_12}
    \end{subfigure}
    \caption{Case studies comparing the dosimetrist's plan to the JulianA plan. The same colorwash is used for both plans, where $100\%$ corresponds to \SI{69.96}{Gy RBE}. The left plots depict the dosimetrist's plan and the right plot the JulianA plan that was generated automatically.}
    \label{fig:case_study}
\end{figure}
We display three case studies in Fig.~\ref{fig:case_study}. One for which the j-plan was considered superior (case 05, Fig.~\ref{fig:case_05}), comparable (case 12, Fig.~\ref{fig:case_12}) and inferior (case 09, Fig.~\ref{fig:case_09}). For case 05, the j-plan achieved a better sparing of the brainstem and the parotids while maintaining comparable target coverage. The dose distributions for case~12 look very similar. The JulianA plan sacrifices more target coverage to spare the right optical structures, but this was not deemed of clinical relevance by the expert reviewers. Case~09 highlights the importance of providing accurate OAR constraints. The prescription handed to JulianA contained a mean-dose constraint for both parotid glands. Therefore, the j-plans aim to spare both parotids. However, it is clinical practice at our institute to spare at least one parotid, with the option to sacrifice the other to improve target coverage. For patient 09, the d-plan sacrifices the left parotid in order to improve target coverage, which was deemed preferable by the expert reviewers.

\subsection{Overall plan quality}
\begin{figure}[ht]
   \begin{center}
   \includegraphics[width=\textwidth]{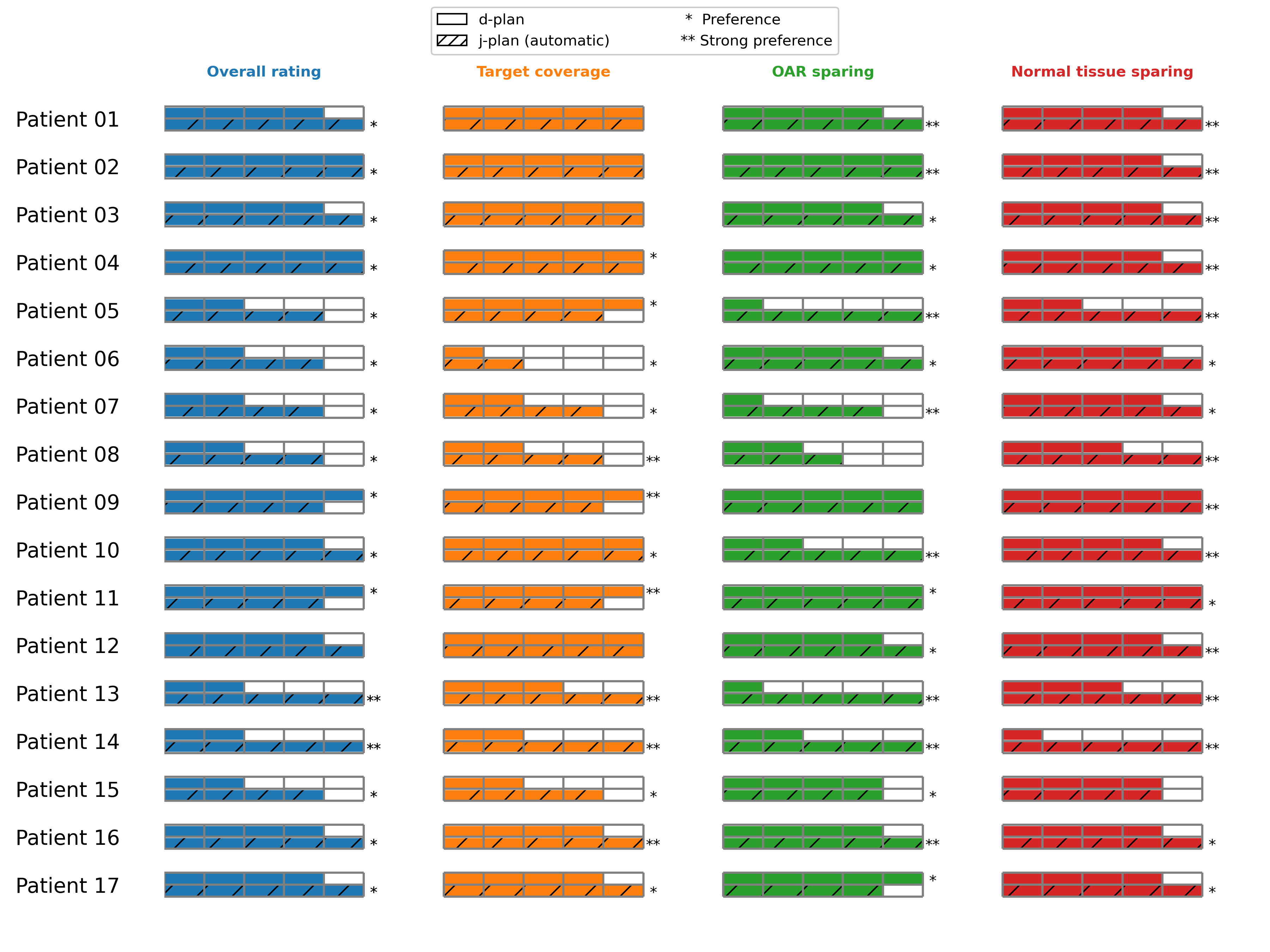}
   \captionv{16}{}
   {Results of the plan review by expert reviewers. Each patient's d- and j-plans were compared w.~r.~t.\ target coverage, compliance with OAR thresholds, normal tissue sparing and overall plan quality. The asterisks indicate preference (*) and strong preference (**).
   \label{fig:MD_review} 
    }  
    \end{center}
\end{figure}

A comprehensive summary of the expert plan assessment by two expert reviewers (JW, AC) is displayed in Fig.~\ref{fig:MD_review}. All the j-plans were deemed acceptable, with a preference for the j-plan in $14 (82.4\%)$~cases, no preference in $1 (5.9\%)$~case and a preference for the d-plan for $2 (11.8\%)$~cases. It is noticeable that $7 (41.2\%)$ of the d-plans were deemed rather unacceptable by the strict standards of this study, which imposed that any violation of an OAR constraint renders a plan unacceptable. However, none of these violations was considered prohibitive in a clinical setting for the following reasons. This study only considers voxel-wise maximum dose and mean dose constraints for OARs. The violated maximum dose constraints were fulfilled when smoother quantities such as $D_{2\%}$ or the $D_{0.03cc}$ were considered (patients 05, 06, 07, 13, 14). Some constraints for small organs like the eyes and the cochleae were violated only according to the \texttt{JulianA.jl} code used in this study, but not according to the Eclipse TPS that was used for the real patient treatment. These disagreement between the systems are caused by ambiguous inside-outside tests when a contour intersects a voxel. They cannot be avoided when evaluating treatment plans generated in different systems. For patient 14, the d-plan sacrificed one cochlea, which is acceptable according to our clinical protocol, but this was not reflected in the prescriptions. Similar arguments apply for the unilateral sparing of the parotids for patient 08 and 09.

There is a general tendency that j-plans are more conformal and achieve sharper dose gradients than their respective d-plan counterparts. This improved conformality is especially striking for Fig.~\ref{fig:case_05}, but can be seen in most of the patients in the cohort. Another commonality is that j-plans tend to overspare the parotids, which can be seen in Fig.~\ref{fig:case_05} \& Fig.~\ref{fig:case_09}. However, this oversparing is mild enough to not compromise the clinical acceptability of the j-plans significantly.

\subsection{Target coverage}
The experts judge the target coverage of the j-plan as superior to the one of the d-plan for $9 (52.9\%)$~cases, equal for $4 (23.5\%)$~cases and inferior for $4 (23.5\%)$ cases. The target coverage of patient~04 is deemed clearly acceptable, which means that the inferiority is not distinct. For patient 05, the d-plan exceeds the maximum dose constraint for the brainstem by \SI{3.0}{Gy RBE}, which improves the target coverage compared to the j-plan that complies with the brainstem constraint. For patient~09, the d-plan opts to exceed the prescription to the left parotid because the right parotid does not receive any dose at all due to the location of the tumour and the choice of beam arrangement. The prescriptions given to JulianA does not include this reasoning. Therefore, JulianA spares both parotids, which deteriorates the target coverage in this case. A similar situation occurs for patient~11, also involving the right parotid.

\begin{figure}[ht]
   \begin{center}
   \includegraphics[width=\textwidth]{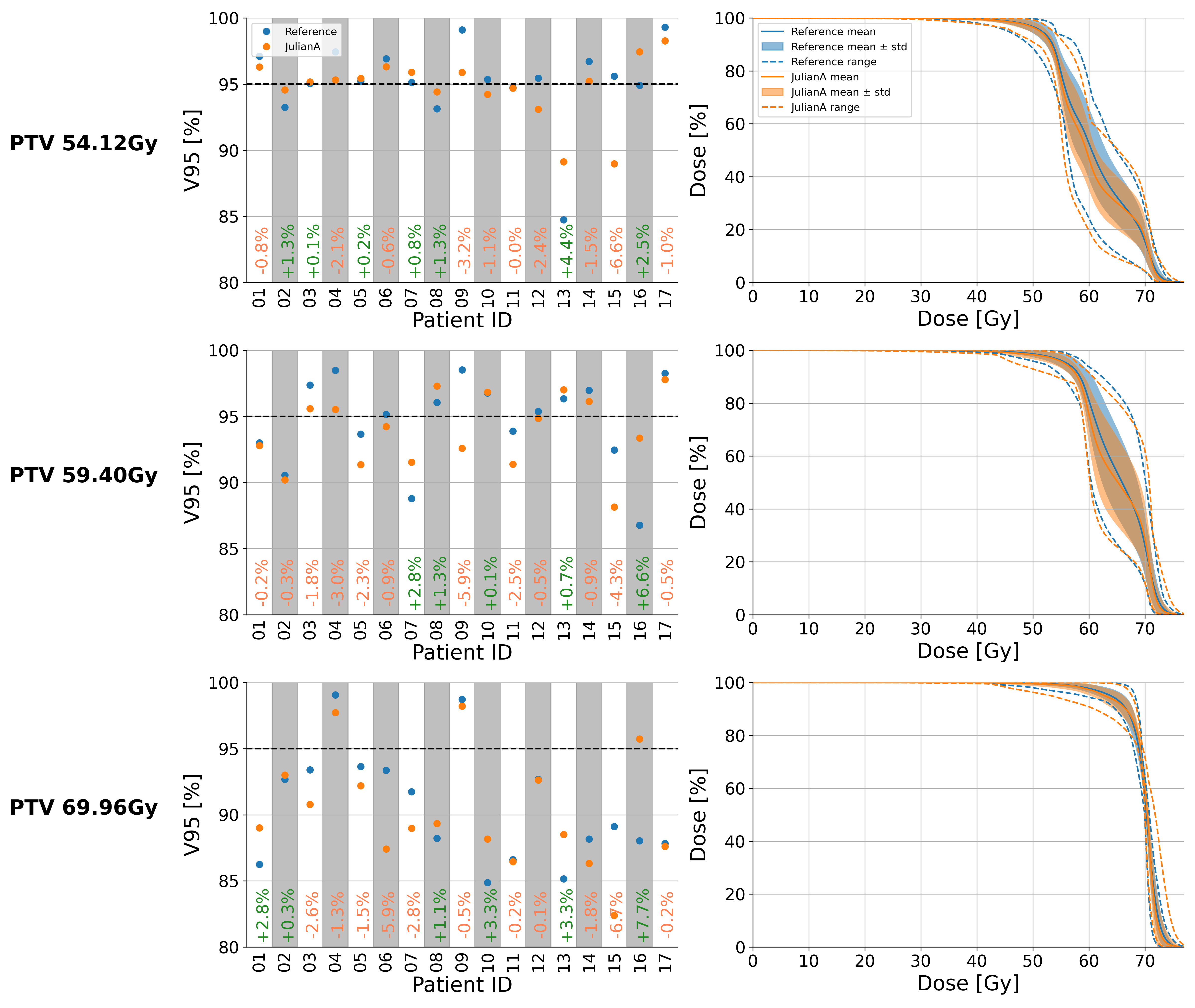}
   \captionv{16}{}
   {Left: Target coverage for each target, measured as $V_{95\%}$ for each patient and target. The dotted lines denote the $95\%$ mark, which is our clinic's acceptability mark. Right: Population dose-volume histograms (DVHs). The solid lines denote mean DVH values, the shaded area adds the standard deviation, and the dashed lines denote the minimum and maximum values, respectively.
   \label{fig:target_coverage} 
    }  
    \end{center}
\end{figure}

The dose-volume histograms (DVHs) for the PTVs (Fig.~\ref{fig:target_coverage}) confirm the results of the expert assessment that JulianA achieves target coverage on par with dosimetrist. Within one standard deviation, the d-plans and the j-plans do not diverge considerably. However, the j-plans show a slight deterioration around the prescribed dose levels when the minimum and maximum curve over the entire population is considered.


The expert evaluation of the target coverage is complemented by an analysis of the $V_{95\%}$, i.~e.\ volume fraction receiving at least $95\%$ of the dose to the corresponding target. Our clinic's acceptability criterion of $V_{95\%} \geq 95\%$ for the lowest-dose planning target volume (PTV) is achieved for $10 (58.8\%)$ cases. For $4 (23.5\%)$ cases, the j-plan does not fulfill the criterion, but shows a better value than the corresponding d-plan. For $3 (17.6\%)$ cases, the j-plan does not fulfill the acceptability criterion and has a lower $V_{95\%}$ value than the d-plan. The j-plan for patient~10, while exhibiting a lower $V_{95\%}$ value, still achieves a more homogeneous and sharper dose distribution than the d-plan. The reduced target coverage in patient~12 originates in oversparing the parotid glands in the j-plan, which is a tendency observed also in other patients (Retina-R violated by d-plan). For patient~15, the d-plan achieves superior target coverage, but does not distinguish between the dose levels, putting as much as \SI{60}{Gy RBE} instead of the prescribed \SI{54.12}{Gy RBE}.

\subsection{Sparing of OARs and healthy tissue}
\begin{figure}[ht]
   \begin{center}
   \includegraphics[width=\textwidth]{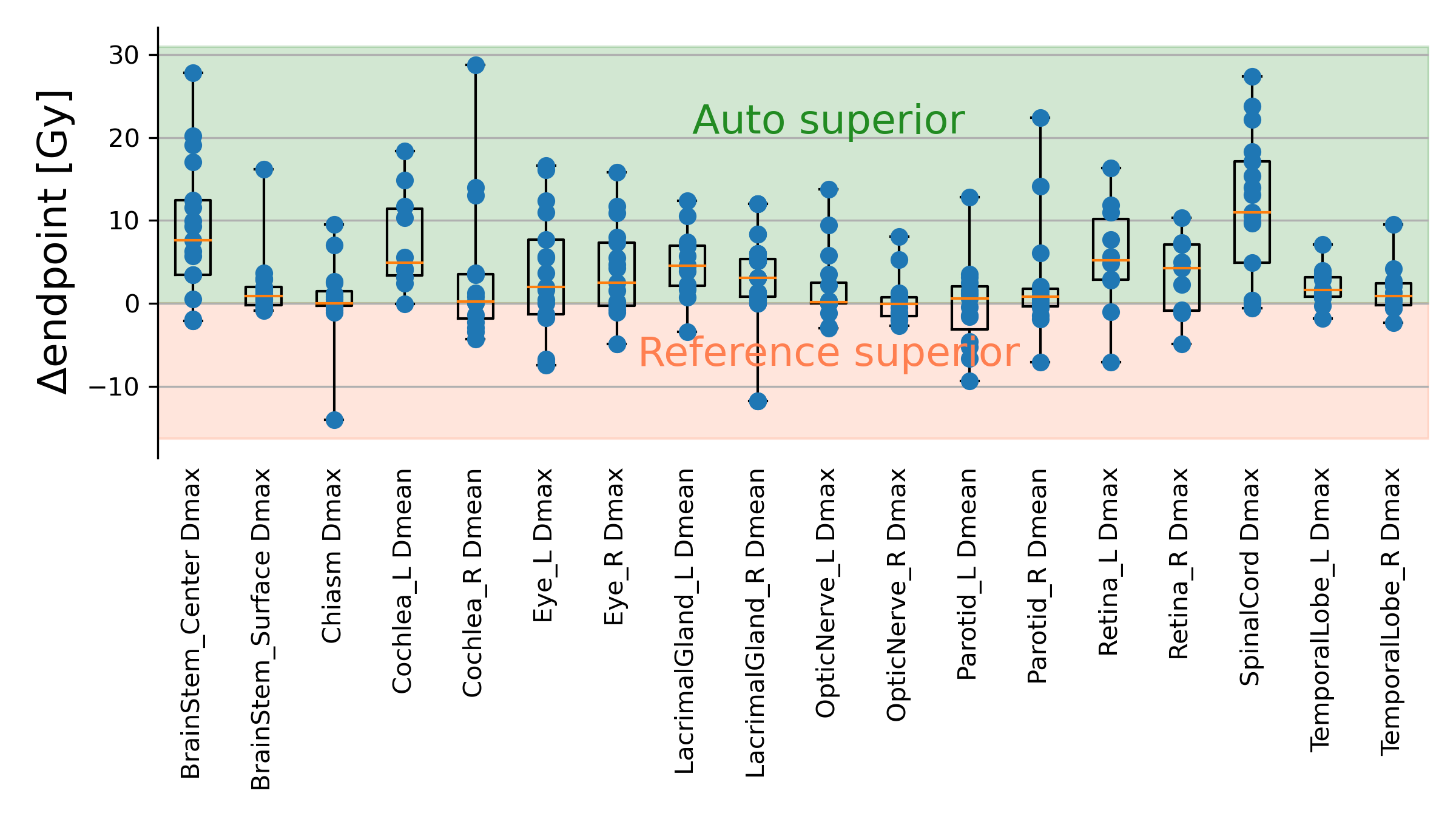}
   \captionv{16}{}
   {Difference in OAR endpoints for each patient (if the structure was contoured). Positive values mean that the d-plan has a higher value, negative values that it has a lower value than the j-plan for the given endpoint. The dots indicate patients and the boxplots summarise the distribution of the dots. The Orange bar is the median difference, the box edges mark the quartiles and the whiskers delimit the range.
   \label{fig:oar} 
    }  
    \end{center}
\end{figure}

The j-plans exhibit superior compliance with OAR constraints for $13 (76.5\%)$, equal compliance for $2 (11.8\%)$ and inferior compliance for $2 (11.8\%)$ cases according to the expert validation. The j-plan for patient~11 spares both parotids, which results in oversparing compared to the d-plan that spares only one side. For patient 17, the j-plan contains a single voxel that exceeds the maximum dose constraint by \SI{0.1}{Gy RBE}, while all voxels of the d-plan remain below the threshold. The normal tissue sparing is considered superior for $16 (94.1\%)$ cases and comparable for $1 (5.9\%)$ of the cases. These results are confirmed by OAR endpoint statistics in Fig.~\ref{fig:oar}. All the median differences of OAR endpoints are favourable for the j-plans, with as high as \SI{10}{Gy RBE} improvement for the maximum dose to the spinal cord compared to the d-plans.

\begin{figure}[ht]
   \begin{center}
   \includegraphics[width=\textwidth]{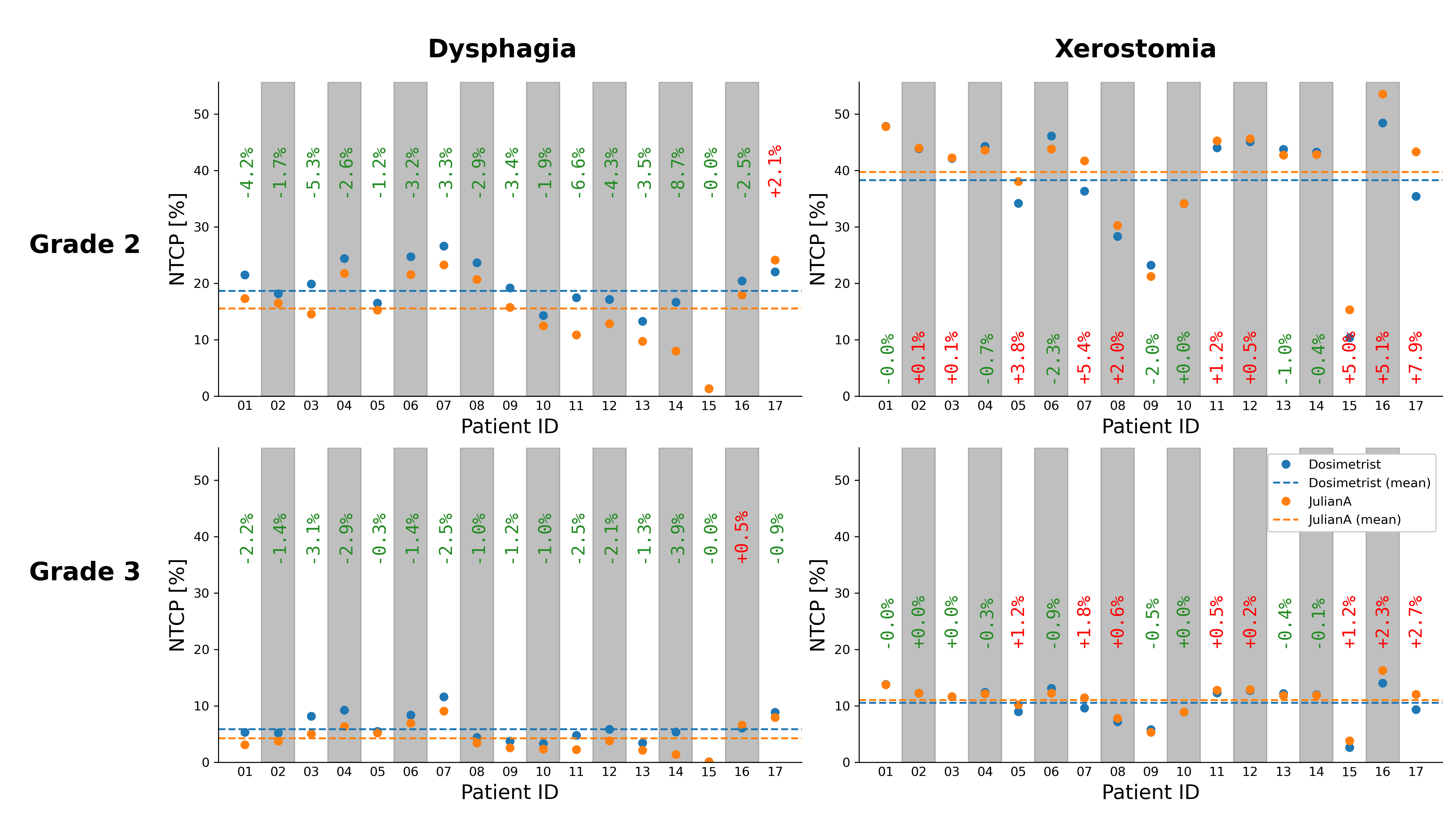}
   \captionv{16}{}
   {Normal tissue complication probabilities for dysphagia (left) and xerostomia (right) for grade two (top) and three (bottom). The j-plans are depicted in red, the d-plans in blue.
   \label{fig:ntcp} 
    }  
    \end{center}
\end{figure}

Fig.~\ref{fig:ntcp} depicts the normal-tissue complication probabilities (NTCPs) of all plans. The mean and standard deviation of the j-plans reduction in probability of dysphagia is $-3.1 \pm 2.4\%$ (range $[-8.7\%, 2.1\%]$) for grade two and $-1.6 \pm 1.2\%$ (range $[-3.9\%, 0.5\%]$ for grade three. For both grades, the j-plans exhibit lower NTCP values for all but one patient. For xerostomia, the corresponding values are $1.5 \pm 2.9\%$ (range $[-2.3\%, 7.9\%]$) for grade two and $0.5 \pm 1.0\%$ (range $[-0.9\%, 2.7\%$) for grade three. Neither the j-plans nor the d-plans show a clear advantage in terms of NTCP.

\subsection{Planning Time}
The average and standard deviation of the time for optimisation was $10.4 \pm 4.4$~min \mbox{(range: 1.6-17.2~min)} on an NVIDIA DGX A100 graphics processing unit (GPU), which is competitive to similar algorithms in literature. The recently published SISS algorithm~\cite{Kong2024} takes \SI{47}{min} for treatment planning and belongs to the category of MCO methods, like JulianA.

\FloatBarrier



\section{Discussion and Conclusion}

JulianA demonstrates its capabilites of creating treatment plans that are comparable to manually created treatment plans, all within the competitive planning times of $2-17$~min. All of the JulianA plans were deemed acceptable. They exhibit superior OAR sparing and lower dose to normal tissue, while maintaining target coverage comparable to the one achieved by human planners for the examined cohort. The expert evaluation yields that the JulianA plans match or surpass the d-plans' OAR sparing in $15 (88.2\%)$~cases and achieve a superior normal tissue sparing for $16 (94.1\%)$~cases and equal normal tissue sparing for the remaining $1 (5.9\%)$ case. Probability of dysphagia was slightly reduced for grade 2 ($-3.1 \pm 2.4\%$) and grade 3 dysphagia ($-1.6 \pm 1.2\%$) compared to the d-plans. For xerostomia probability, no significant difference could be observed ($1.5 \pm 2.9\%$ for grade 2 and $0.5 \pm 1.0\%$ for grade 3). Therefore, JulianA plans exhibit similar NTCP values as the d-plans.\\
The improved OAR and normal tissue sparing comes at the cost of slightly reduced target coverage. Nevertheless, the expert evaluation has shown that JulianA's target coverage was only inferior for $4$~cases and rendered none of them unacceptable. We therefore conclude that JulianA selects a different tradeoff between target coverage and OAR sparing than the human planners and that the JulianA tradeoff leads to plans of equal or even better quality for $15 (88.2\%)$ of the cases in this dataset. The remaining $2 (11.8\%)$ cases were considered inferior mainly due to oversparing both parotids instead of following our clinical practice of sparing only one parotid. This oversparing could be mitigated by adjusting the prescription to constrain only one parotid. Another, more flexible option would be to manually fine-tuning the default objective importance of the parotid objectives. Fine-tuning is possible for all objectives, which allows for manual adjustments in order to apply JulianA to especially difficult non-standard cases.

The results in this study demonstrate that the current version of JulianA is ready for several applications in our institute. First, the JulianA plans could support the tumour board by providing a first impression of the challenges associated with a given patient. The dosimetrists could then specifically invest more time to improve the distribution in challenging regions and use the plan as a reference to quickly assess intermediate results while developing the plan for treatment. The JulianA plans could also be used for automatic plan quality checks. Second, JulianA could play a crucial role in patient selection similar to the Dutch model-based approach~\cite{Langendijk2013}. In such a scenario, it would replace a KBP tool even in scenarios where no training dataset is available. Third, the JulianA plans could be used to train aspiring dosimetrists and medical physicists. Finally, JulianA is suitable to support researchers. We have already applied the JulianA autoplanning to create treatment plans for $12$ patients within a single day in order to validate a machine learning model \cite{Li2024-ty}. The strength of an autoplanning tool are that the entire training and validation pipelines can be run automatically. Additionally, domain experts such as computer scientists are relieved from the need to learn treatment planning and the burden to create a large number of treatment plans manually. All these applications can be realised using the open source Julia package JulianA.jl~\cite{bellotti_2024_11079457}.

Despite these encouraging results, some challenges need to be addressed before JulianA can be used for patient treatment. First, JulianA shows a tendency of oversparing the parotids. This tendency is not distinct, but should be mentioned for the sake of completeness. Second, the JulianA plans utilise a high number of spots compared to the Eclipse plans. The reason for this is that currently, JulianA relies on our in-house TPS FIonA for the spot placement. There is ongoing research at our institute to reduce the number of spots. Third, the JulianA plans in this study were optimised non-robustly because the historic reference plans were also optimised non-robustly. On the one hand, we recognise that robustness is crucial for proton therapy. On the other hand, our institute introduced robust optimisation into our workflows only recently. Therefore, this study aims to demonstrate JulianA's utility for the bulk of patients treated at our institute so far. To the best of our knowledge, there is no fundamental limit in the introduction of robustness since robustness mainly requires the calculation of multiple dose distributions for various uncertainty scenarios in each iteration of the optimisation.

A promising application of JulianA could be in the emerging field of daily-adaptive proton therapy (DAPT), where it would reoptimise the plan of the day. However, DAPT requires short runtimes for all steps in the workflow. Currently, JulianA is too slow for such a workflow. However, advances in computation power of GPUs might accelerate the code enough for DAPT.

Future research will aim to validate JulianA for sites other than head-and-neck cases. Further, robust optimisation must be introduced in future versions of JulianA. Finally, the number of spots needs to be reduced to a level comparable to commercial systems. Once these improvements are in place, we are confident that JulianA can be used to treat patients at our institute.

In summary, we have shown that JulianA is capable of optimising spot weights within $2-17$~min without any human interaction or patient-specific adjustments. All of the automatically generated plans are acceptable for treatment at our institute according to the two expert reviewers. While further improvements are needed for routine patient treatment using JulianA, these results demonstrate that JulianA is ready to support clinical operations as a QA tool and improve research by reducing time spent on treatment planning and making high-quality treatment plans available to researchers.



\newpage     

\section*{Appendix}
\addcontentsline{toc}{section}{\numberline{}Appendix}
\subsection*{Code availability}
The code used in this study is published on Zenodo \cite{code}.

\section*{References}
\addcontentsline{toc}{section}{\numberline{}References}
\vspace*{-20mm}





\bibliography{./juliana_hn}      



\bibliographystyle{medphy.bst}    


\end{document}